\documentstyle[aps,manuscript]{revtex}

\begin{document}

\title {An Improved Quantum Molecular Dynamics Model and its
Applications to Fusion Reaction near Barrier
\footnote {Supported
by National Natural Science Foundation of China, No. 19975073,
10175093, 10175089 and Science Foundation of Chinese Nuclear
Industry and Major State Basic Research Development Program and
Contract, No. G20000774.
\newline
 E-Mail: wangning@iris.ciae.ac.cn(Ning Wang),lizwux@iris.ciae.ac.cn(Zhuxia Li and Xizhen Wu)}}

\author {Ning Wang$^{1}$, Zhuxia Li$^{1,2,3}$, Xizhen Wu$^{1,2}$}

\address { 1) China Institute of Atomic Energy, P. O. Box 275(18), Beijing 102413, P. R. China\\
           2) Nuclear Theory Center of National Laboratory of Heavy Ion Accelerator, Lanzhou P. R. China\\
           3) Institute of Theoretical Physics, Chinese Academic of Science, Beijing 100080  P. R. China }
\maketitle

\begin{abstract}
\sf An improved Quantum Molecular Dynamics model is proposed. By
using this model, the properties of ground state of nuclei from
$^{6}$Li to $^{208}$Pb can be described very well with one set of
parameters. The fusion reactions for $^{40}$Ca+$^{90}$Zr ,
$^{40}$Ca+$^{96}$Zr and $^{48}$Ca+$^{90}$Zr at energy near barrier
are studied by this model. The experimental data of the fusion
cross sections for $^{40}$Ca+$^{90,96}$Zr at the energy near
barrier can be reproduced remarkably well without introducing any
new parameters. The mechanism for the enhancement of fusion
probability for fusion reactions with neutron-rich projectile or
target is analyzed.
\\
\end{abstract}

\begin{center}
{\bf 1. INTRODUCTION}
\end{center}

Heavy ion fusion reactions at energies below and near the Coulomb
barrier have received considerable attention. The dynamical
mechanism of fusion reactions such as the dynamical process of
target and projectile deformation before contact and the neck
formation and development after contact are of special interest.
The Quantum Molecular Dynamics model (QMD) is a microscopic
dynamical model and can successfully provide quite a lot of
dynamic information about reaction mechanism (for example
see\cite{Hart89,Ai91,Hart98}). However, it is still of
difficulties to apply QMD to low energy reactions. The main
difficulty for this case is that one has to deal with a problem of
time evolution of nuclear many body systems which are of Fermionic
nature. On quantum mechanical level, the wave function must be
antisymmetrized because of the Fermionic nature of the nuclear
constituents, such as did in Antisymmetrized Molecular
Dynamics(AMD)\cite{amd} and Fermionic Moleculare Dynamics
(FMD)\cite{Fel97}. AMD and FMD has made a great achievement on
describing the nuclear reaction and structure for light nuclei.
Nevertheless, for AMD and FMD one has to deal with N! components
to the wave function even for a single slater determinant.
Therefore the time evolution by numerically propagating the
nuclear wave function would be computationally very demanding and
the CPU time necessary to work out calculations for systems with
total mass larger than 200 is very large for practical
studies\cite{Papa00}. So one is forced to make a compromise
between approximating the many body physics and producing a
simulation which can execute fast enough to do practical studies.
In QMD model, each nucleon is represented by a coherent state of
the form $\phi_{i}$ and the total N-body wave function is assumed
to be the direct product of coherent states, $\Phi=\Pi\phi_{i}$.
So QMD calculation is much easer and faster to be carried out. Of
cause, it is clear that QMD model does not use a slater
determinant and thus antisymmetrization is neglected, although the
two-body collisions with Pauli blocking have some effects to
maintain a part of Fermionic feature of systems. In fact, two-body
collisions are very rare in ground states or in fusion reactions.
To compensate this shortcoming, the two-body Pauli potential is
introduced by several authors \cite{David,Wilet77,Dorso87,Maru98}
to mimic the Pauli principle. Although the Pauli potential can
improve the ground states to a certain extent, it is not effective
enough to avoid from evolving the initial phase space distribution
obtained from sampling the nuclear ground state density to a
classical Boltzmann one after a long time\cite{Papa00}.

 In addition, the width of Gaussian wave packet in QMD model is taken to be a
constant but we find the values adopted are quite different for different
calculations. For example,in \cite{Hart98} the Gaussian wave packet width
was taken as $L=4.33fm^{2}$ for reaction Ca+Ca and $L=8.66fm^{2}$ for Au+Au.
And in ref.\cite{Toshi01}, the authors took
two different values of the width of Gaussian wave packet for
multifragmentation and fusion reaction. Therefore it seems to us that it
is worthwhile to make further study on the influence of the width of
 Gaussian wave packet in QMD model calculation.

Aiming at studying the dynamical process of fusion reaction at the energy near
barrier one needs quite stable target and projectile nuclei which
can bound nucleons for long enough time without spurious particle
emission since fusion reaction processes usually last long time.
It seems to be difficult for the normal QMD model and an improved QMD model is
required for this purpose.

Based on above discussions we develop an improved QMD model in this work.
With this model, not only the bulk properties of the ground state of nuclei including density and
momentum distribution, binding energy, mean square radius etc. but also
their time evolution can be described correctly.
And simultaneously we apply it to describing fusion reactions at the energy near
barrier and a good agreement with experiment results is achieved.

The structure of the article is as follows: In the section 2, we
introduce our improved QMD model. Then we make application of
this model to describing the nuclear ground states and the fusion
reaction process. The results are presented in section 3. Finally a short
summary and conclusion is given in section 4.

\begin{center}
\bigskip {\bf 2. THE IMPROVED QMD MODEL}
\end{center}

In this section we introduce the improved QMD model in more
details. First, a brief introduction to QMD model is presented.
Then, the main improvements are introduced and their effects
are analyzed. Finally, the preparation procedure of initial nuclei is
given.

\begin{center}
{\bf 2.1 The Brief Introduction}
\end{center}

In QMD, each nucleon is represented by a coherent state of a
Gaussian wave packet,
\begin{equation}  \label{1}
\phi _{i}({\bf r})=\frac{1}{(2\pi \sigma _{r}^{2})^{3/4}}\exp [-\frac{%
({\bf r-r}_{i})^{2}}{4\sigma _{r}^{2}}+\frac{i}{\hbar}{\bf r}\cdot
{\bf p}_{i}],
\end{equation}
Where, ${\bf r}_{i}, {\bf p}_{i}$, are the centers of i-th wave packet in the
coordinate and momentum space, respectively. $\sigma _{r}$
represents the spatial spread of the wave packet. The total N-body
wave function is assumed to be the direct product of these
coherent states. Through Wigner transformation of the wave function,
the N-body phase space distribution function is given by:
\begin{equation}  \label{2}
f_{i}({\bf
r,p})=\sum\limits_{i}\frac{1}{(\pi\hbar)^{3}}\exp[-\frac{({\bf
r-r}_{i})^{2}}{2\sigma_{r}^{2}}-\frac{2\sigma_{r}^{2}}{\hbar^{2}}({\bf
p-p}_{i})^{2}].
\end{equation}

The density and momentum distribution of a system respectively
reads as:
\begin{equation}  \label{3}
\rho ({\bf r})=\int f({\bf r,p})d^{3}p=\sum\limits_{i} \rho
_{i}({\bf r}),
\end{equation}

\begin{equation}  \label{4}
g({\bf p})=\int f({\bf r,p})d^{3}r=\sum\limits_{i} g_{i}({\bf p}),
\end{equation}
where the sum runs over all particles in the system, and $\rho
_{i}({\bf r})$ and $g_{i}({\bf p})$ are the density and momentum
distribution of nucleon i:
\begin{equation}  \label{5}
\rho _{i}({\bf r})=\frac{1}{(2\pi \sigma _{r}^{2})^{3/2}}\exp [-\frac{%
({\bf r-r}_{i})^{2}}{2\sigma _{r}^{2}}],
\end{equation}

\begin{equation}  \label{6}
g_{i}({\bf p})=\frac{1}{(2\pi \sigma _{p}^{2})^{3/2}}\exp [-\frac{%
({\bf p-p}_{i})^{2}}{2\sigma _{p}^{2}}],
\end{equation}
where $\sigma_{r}$ and $\sigma_{p}$ are the widths of wave packets in
coordinate and momentum space, respectively and they satisfy the
minimum uncertainty relation:
\begin{equation}  \label{7}
\sigma_{r}\sigma_{p}=\frac{\hbar}{2}.
\end{equation}

In QMD, the nucleons in a system move under the selfconsistently
generated mean-field, and the time evolution of ${\bf r}_{i},{\bf p}_{i}$ is
governed by Hamiltonian equation of motion:
\begin{equation}  \label{8}
\dot{{\bf p}}_{i}=-\frac{\partial H}{\partial {\bf r}_{i}}, \dot{{\bf r}}_{i}=\frac{\partial H}{%
\partial {\bf p}_{i}}.
\end{equation}

The Hamiltonian H consists of the kinetic energy and effective
interaction potential energy,
\begin{equation}  \label{9}
H=T+U,
\end{equation}
\begin{equation}  \label{10}
T=\sum\limits_{i} \frac{{\bf p}_{i}^{2}}{2m}.
\end{equation}

The effective interaction potential energy includes the nuclear
local interaction potential energy and Coulomb interaction
potential energy,
\begin{equation}  \label{11}
U=U_{loc}+U_{coul}.
\end{equation}

And
\begin{equation}  \label{12}
U_{loc}=\int V_{loc}d^{3}{\bf r},
\end{equation}
$V_{loc}$ is the potential energy density and can be derived
directly from a zero-range Skyrme interaction\cite{David,Vau72}.
\begin{equation}  \label{13}
V_{loc}=\frac{\alpha }{2}\frac{\rho ({\bf r})^{2}}{\rho _{0}}+\frac{\beta }{3}%
\frac{\rho ({\bf r})^{3}}{\rho
_{0}^{2}}+\frac{C_{s}}{2}\frac{(\rho _{p}({\bf r})-\rho _{n}({\bf
r}))^{2}}{\rho _{0}}+\frac{g_{1}}{2}(\nabla \rho ({\bf r}))^{2}.
\end{equation}
By using
\begin{equation}  \label{14}
\langle Q\rangle _{i}=\int \rho _{i}({\bf r})Qd^{3}{\bf r},
\end{equation}
the nuclear local interaction potential energy can be written as:
\begin{equation}  \label{15}
U_{loc}=\frac{\alpha }{2}\sum\limits_{i} \langle \frac{\rho }{\rho _{0}}\rangle _{i}+%
\frac{\beta }{3}\sum\limits_{i} \langle \frac{\rho ^{2}}{\rho _{0}^{2}}\rangle _{i}+%
\frac{C_{s}}{2}\int \frac{(\rho _{p}-\rho _{n})^{2}}{\rho
_{0}}d^{3}{\bf r}+\int \frac{g_{1}}{2}(\nabla \rho )^{2}d^{3}{\bf r}.
\end{equation}

Because of the Gaussian form of density distributions in Eq.(5),
all of the
integrals in Eq.(12) can be done analytically, furthermore all but
one of the sums involves only $N^{2}$ terms. The problem in
Eq.(15) is that $\sum\limits_{i}\langle \frac{\rho ^{2}}{\rho
_{0}^{2}}\rangle _{i}$ is of order of $N^{3}$ and for a system of
hundreds of particles, evaluation of $N^{3}$ elements is very
time-consuming and computationally prohibitive, so it is
approximated by\cite{David}
\begin{equation}  \label{16}
\sum\limits_{i} \langle \frac{\rho ^{2}}{\rho _{0}^{2}}\rangle
_{i}\approx \sum\limits_{i} \langle \frac{\rho }{\rho _{0}}\rangle
_{i}^{2}+\int \frac{g_{2}}{2}(\nabla \rho )^{2}d^{3}{\bf r},
\end{equation}
which is a $N^{2}$ operation.\ Since the second term in Eq.(16)
has the same functional form as the surface energy term in
Eq.(15), we combine them into one term and call it the surface energy
term with parameter $g_{0}=g_{1}+g_{2}$.
The Coulomb potential energy is obtained from:
\begin{equation} \label{17}
U_{coul}=\frac{1}{2}\sum\limits_{i\neq j} \int \rho _{i}({\bf
r})\frac{e^{2}}{|{\bf r-r}^{\prime }|}\rho _{j}({\bf r}^{\prime
})d^{3}{\bf r}d^{3}{\bf r}^{\prime }.
\end{equation}

The parameters in this work are listed in Table.1.

\[
\fbox{Table. 1 }
\]

\begin{center}
{\bf 2.2 The Effect of Surface Energy Term and Phase Space Constraint}
\end{center}

It is obvious that the surface effects are important for a finite
system. Let us first study the effects of the term
$U_{surface}=\frac{g_{0}}{2}\int (\nabla \rho )^{2}d^{3}r$. In
Fig.1, we show a schematic figure of the effect of the surface
energy term. As mentioned in the introduction, the initial density
distribution of a system will evolve to a classical one according
to classical equations of motion after long enough time. Suppose
we have a Gaussian form of density distribution as shown in
Fig.1(a). With this density distribution, the surface energy term
$U_{surface}$ is obtained by definition, and its shape is shown in
Fig.1(b). From the figure we can see that the particles in the
central region will experience  a repulsion and are forced to move
toward outside. Consequently, the density at central region is
suppressed and avoids going up unreasonable high. While the
particles at surface feel an attraction and move toward inside so
that the density distribution at the surface will not extend too
far. Fig.1(c) shows the influence of the surface energy term on
the density distribution of the system. The solid curve and dashed
curve are the density distribution calculated with and without the
surface term taken into account, respectively. It is clear that
the density distribution calculated with the surface term is more
reasonable than that without the surface term.

\[
\fbox{Fig. 1 }
\]

As is well known, the Guassian wave packet itself has a long tail
which makes the surface of the system more disperse. Therefore, in
QMD model it is more important to introduce the surface energy
term. Here we pay great attention to the surface energy term, and
make further study about the effect of the surface term for
realistic cases.  In Fig.2 and Fig.3 we show the time evolution of
the density distribution of $^{90}$Zr calculated without and with
surface term taken into account. The initial density distribution
is obtained by relativistic mean field theory (RMF)
calculation\cite{Rein86}. When the surface term is not included,
as Fig.2 shows, the density distribution keep stable and the
central density of nuclei maintains lower than $0.2fm^{-3}$ only
at the early stage(for example, see $t=5fm/c$)for all runs. With
further time evolution, the density distribution changes and
deviates from the initial one. After about $t=200fm/c$, the
central density begins to go up, and at about $t=400fm/c$ the
central density even reaches $0.3fm^{-3}$. And after $t=400fm/c$
there is spurious emission of nucleons while the central density
is still too high. When the surface term is included, as Fig.3
shows, with time evolution the shape of density distribution can
keep remarkably stable as the same as the initial one, and the
central density always keeps at $\rho=0.165fm^{-3}$. Even at
$t=800fm/c$, the central density still remains the same value of
$\rho=0.165fm^{-3}$. From the comparison of these two figures, one
can clearly see that the surface term is effective to maintain a
reasonable density distribution for a ground state of nucleus
during time evolution.

\[
\fbox{Fig. 2 }
\]
\[
\fbox{Fig. 3 }
\]

Similarly, the momentum distribution will evolve to a classical
distribution from initial momentum distribution after long time.
To avoid it, a phase space density constraint was introduced by
Papa et. al. in ref.\cite{Papa00}. Because Fermionic nature
requires one-body phase space density of a system \={f}$_{i}\leq
1$, in this work we perform many body elastic scattering to reduce
the phase space occupation if the phase space occupation
\={f}$_{i}$ is greater than 1, as did in ref.\cite{Papa00}. At the
same time the Pauli blocking probability is checked like the usual
treatment in two-body collision process. This kind of constraint
affects the low momentum part of the momentum distribution
strongly and can effectively avoid the number of particles with
low momentum becoming too large. Fig.4 shows the comparison of the
time evolution of the average momentum distributions calculated
without taking constraint (dashed curves) and with taking
constraint (solid curves) for 200 $^{208}$Pb nuclei. Here the
momentum distribution means the distribution of the centroid
momentums of wave packets of nucleons in a system. From the figure
one can see that at the initial time, the momentum distribution
(dash dotted curves) is reasonable. With time evolution, the
difference in the momentum distributions for two cases becomes
obvious. When the phase space density constraint is not taken into
account, the number of particles with low momentum increases
greatly and the momentum distribution deviates from the initial
one clearly, as shown in dashed curves. This problem is usually
ignored in the QMD calculation when it is applied to medium or
high energy reactions. However for describing ground state or
fusion reactions near barrier, this problem should be considered
seriously. From solid curves one can find the behavior of time
evolution of momentum distribution,especially the low momentum
part is improved a lot after taking the phase space density
constraint into account. However we notice that the high momentum
part of the distribution is still too disperse comparing with the
initial one. This means that phase space constraint is still not
enough to control the momentum distribution as good as request.
But even so we find that it improves the fusion reaction near
barrier a lot. It may be because the high momentum part is not so
important for this case. The investigating on further improvement
of the behavior of the time evolution of momentum distribution is
in progress.

\[
\fbox{Fig. 4 }
\]

\begin{center}
{\bf 2.3 The Width of Wave Packet}
\end{center}

In QMD, the width of wave packet can be regarded as a quantity
having relations with the interaction range of a particle. Its
influence disappears for infinite nuclear matter whereas for
finite systems it may play a un-negligible role\cite{Hart98}. In
the normal QMD model, the width of wave packet is taken as a constant.
For example, in ref.\cite{Hart981}, the width of wave packet is
taken as $\sigma _{r}=1.04fm$ for Ca+Ca, and $\sigma _{r}=1.47fm$
(i.e. $L=8.66fm^{2}$ in notation of \cite{Hart981}) for Au+Au.
Here let us study the influence of the width of wave packet on the
stability of a nuclear system. As an example, for the ground state of $^{208}$Pb,
if we choose $\sigma_{r}=1.04fm $, there are about 30 spurious particles emitted
until 800fm/c while if $\sigma _{r}=1.44fm$,
 there are no particles emitted until 800fm/c generally and simultaneously both
  the density and momentum distribution are reasonable.
  In heavy ion reactions, if spurious emission becomes
serious, the result will be affected obviously. Therefore in the
study of reaction of Au+Au in ref.\cite{Hart981}, a larger width of
wave packet was taken, i.e. $\sigma _{r}=1.47fm$, and we find it
is reasonable and necessary. In fusion reaction at energy near
barrier, the reaction process lasts a long time, therefore in
QMD calculations, to make nuclei stable enough is vital to the
result for fusion reactions. On the other hand, for Ca or lighter nuclei, if
a large width of wave packet is adopted, the
fluctuation of mean square root of radius of the system becomes too large
during time evolution, which is obviously not suitable.
Therefore for Ca+Ca system, for example in ref.\cite{Hart981} a
small wave packet width such as $\sigma _{r}=1.04fm$ was adopted.
In addition, we also find that it is very important to take a reasonable value
of wave packet width in order to describe the
Coulomb barrier correctly. We will show it in section 3.
Based on above discussion we propose
a system size dependent wave packet width. In our present work, we
take the dependence of the wave packet width on system size as:
\begin{equation}\label{18}
\sigma_{r}=0.16N_{A}^{1/3}+0.49,
\end{equation}
where $N_{A}$ is the number of nucleons binding in a system A
which can be a individual nucleus, or clusters or a compound
nucleus produced in heavy ion reactions according to the specific
problem studied. After introducing the system size dependent wave
packet width, our model is expected be able to well describe the
bulk properties of nuclei in a wide mass region from $^{6}$Li to
$^{208}$Pb as well as the behavior of fusion process.

\begin{center}
{\bf 2.4 Preparation of Initial Nuclei}
\end{center}

As is well known, the initial condition is very important in QMD
calculations. In the present work, the preparation of initial
nuclei is as follow:

Firstly, the neutron and proton density distributions of nuclei
are obtained by means of RMF calculations. Then the position of each nucleon in
nuclei is sampled according to the density distribution obtained.

Secondly, based on the density distribution obtained above,
the Fermi momentum $P_{F}$ is calculated by the local
density approximation. Considering the momentum of each nucleon
being also represented by a wave packet with a width of
$\sigma_{p}$ which satisfies the minimum uncertainty relation
$\sigma_{r}\sigma_{p}=\frac{\hbar}{2}$, the Fermi momentum
$P_{F}^{\prime }$ adopted in the sampling of the momentums of nucleons
should be smaller than $P_{F}$. Here we take the difference
$\Delta P_{F}=P_{F}-P_{F}^{\prime }$ to be about the width of half height of
Guassian wave packet in momentum space (see expression(6)). For
some light nuclei (mass is less than 16) we make a slight
adjustment of the difference $\Delta P_{F}$ which is less than tenth of
$\Delta P_{F}$. Thus we can prepare the initial nuclei from
$^{6}$Li to $^{208}$Pb of which the binding energies and mean
square radius are in good agreement with experimental data.

It is important to have stable initial nuclei with no spurious particle emission.
To check the stability of the pre-prepared initial
nuclei, we let the pre-prepared nuclear systems evolve for at
least 600fm/c, then the ground state properties including mean
square radius, binding energy, density distribution, momentum
distribution, phase space distribution, etc. are checked elaborately.
Only those pre-prepared nuclei for which the bulk properties are good
enough, the behavior of the time evolution of all these
properties maintain stable and there is no spurious particle
emission are selected as 'good initial nuclei' and are stored
for usage in simulating reactions.

\begin{center}
{\bf 3. RESULTS}
\end{center}
\begin{center}
{\bf 3.1 Properties of Ground State}
\end{center}

In Table.2 we give the calculated results of binding energy and mean
square root of radii for
$^{6}$Li,$^{16}$O,$^{30}$P,$^{40}$Ca,$^{90}$Zr,$^{108}$Ag,$^{144}$Nd,$^{197}$Au
and $^{208}$Pb. The binding energies are compared
with experimental data and the mean square roots of radii are compared with
the empirical formula\cite{Pre75}
\begin{equation}  \label{19}
\langle r^{2}\rangle ^{1/2}=0.82A^{1/3}+0.58.
\end{equation}

\[
\fbox{Table. 2}
\]

One can see that the calculated binding energies are in good
agreement with experimental data and the mean square roots of
radii are also in good agreement with empirical values obtained
from the empirical formula(19) except for small nuclei. For small
nuclei, our results are a little bit better than the empirical
formula when we compare them with experimental data(see ref.12).
Considering how few parameters we use in this model, the obtained
results in describing the ground state properties of nucleus are
quite satisfied. In addition to fulfilling the static properties
of ground state of nuclei, such as the binding energies, the mean
square roots of radii, the behavior of the time evolution of those
quantities are also very concerned. In Fig.5 we show the time
evolution of the binding energies and mean square roots of radii
 of $^{16}$O, $^{40}$Ca,$^{90}$Zr and $^{208}$Pb.
One can see that for these nuclei the binding energy and mean
square roots of radii can maintain stable for at least 600fm/c.
One can also find that the larger the size of nucleus is, the
smaller the fluctuation of binding energy and mean square root of
radius with time evolution is, which is because the mean field
effect becomes stronger as system size increases.

\[
\fbox{Fig. 5 }
\]

\begin{center}
{\bf 3.2 Coulomb Barrier}
\end{center}

Coulomb barrier is a very important quantity in
describing fusion reactions. It's height and width are two
sensitive parameters in WKB calculations of fusion cross
sections. The Coulomb barrier can be calculated microscopically
using the following expression in QMD,
\begin{equation}  \label{20}
V_{b}(r)=\int d^{3}r_{1}\int d^{3}r_{2}\rho _{1}({\bf r}_{1}-{\bf
r}_{1c})V({\bf r}_{1}-{\bf r}_{2})\rho _{2}({\bf r}_{2}-{\bf
r}_{2c}),
\end{equation}
where $\rho _{1}$, $\rho _{2}$ are the density distribution of
projectile and target nuclei, respectively, ${\bf r}_{1c}$, ${\bf
r}_{2c}$ are their centers of mass, respectively. $r=|{\bf
r}_{1c}-{\bf r}_{2c}|$ is the distance of projectile and target
nuclei. In QMD both static and dynamic Coulomb barrier
can be calculated.  For calculating the static
Coulomb barrier of fusion reaction, the static density
distributions of both projectile and target are taken. In the
present work, we first let the initial projectile and target
nuclei evolve individually under their self-consistent mean field for about
300fm/c. Then, we take the density distribution at
this time as the static density distribution to calculate static
Coulomb barrier. The dynamical Coulomb barrier is the barrier
experienced during the fusion process and is calculated based on
the instantaneous density distribution of the system in the real
reaction process.

Fig.6 shows the average static Coulomb barrier of $^{40}$Ca+$^{90}$Zr
fusion reactions. The solid curve denotes the result of our improved
QMD model, the crossed curve denotes the result of
proximity potential\cite{Myers00}. One can see that two results by
using two quite different models agree with each other remarkably
well except for the case when two nuclei overlapping happens. The reason for this
deviation is that the proximity potential is only applicable to the case when two nuclei do not
overlap and so as soon as the projectile and target overlap it may not be able to give accurate result
any more. While QMD model is a microscopic model and can
be applicable to both cases without and with overlapping of two nuclei.
\[
\fbox{Fig. 6 }
\]

In order to study the effect of the system size dependence of wave packet
width on Coulomb barrier, in Fig.7 we show the static Coulomb barrier for
$^{40}$Ca+$^{90}$Zr system calculated by fixed wave
packet widths and the system size dependent wave packet width.
The dashed curve denotes the calculated results with
fixed wave packet width $\sigma_{r}=1.3fm$\cite{Papa00} and the
solid curve denotes the results with system size dependent wave
packet width given by expression(18). For $\sigma _{r}=1.3fm$
case, the Coulomb barrier calculated by QMD is lower than that of
the case with size dependent wave packet width. In this figure, we
find that the wave packet width of nucleon affects the density
distribution of the system and thus affects the Coulomb
barrier considerably. Fig.6 and Fig.7 show us that it is important to introduce the
system size dependent wave packet width for describing
Coulomb barrier correctly.
\[
\fbox{Fig. 7 }
\]

\begin{center}
{\bf 3.3 Fusion Reaction}
\end{center}

After performing the procedure of preparation of initial nuclei
as mentioned in the previous section, from thousands of pre-prepared systems,
we elaborately select 10 projectile nuclei and 10 target nuclei.
By rotating these prepared projectile and target nuclei around their centers of mass
by an Euler angle chosen randomly, we create 100 bombarding events
for each reaction energy $E$ and impact parameter $b$.
Through counting the number of fusion events, we obtain
the probability of fusion reaction $g_{fus}(E,b)$, then the cross
section is calculated by using the expression:
\begin{equation}  \label{21}
\sigma _{fus}=2\pi \int\limits_{0}^{b_{\max }}bg_{fus}(E,b)db=2\pi
\sum bg_{fus}(E,b)\Delta b.
\end{equation}
The distance from projectile to target at initial time is taken to be $%
l=20fm.$

The definition of fusion in QMD model is still a difficult problem
which needs to consider carefully. In TDHF calculations, the
fusion event is defined rather operationally as the event in which
the coalesced one-body density survives through one or more
rotations of composite system or through several oscillations of
its radius. In this work we also use the same definition of fusion
event with that in TDHF calculations. In addition, considering the
specific feature of QMD calculations, if the event in which one or
several nucleons escape prior to the formation of compound nucleus
is still belong to fusion event\cite{Maru98}.  Here, for an event if the number
of nucleons escaped during the process of forming compound nuclei
is equal or less than 6, we consider the event as fusion
event.

Fig.8 shows the fusion cross sections for
(a) $^{40}$Ca+$^{90}$Zr and (b) $^{40}$Ca+$^{96}$Zr, respectively.
Experimental data are taken from ref.\cite{Timm98}.
One can see that the calculation results with our improved QMD
model agree with the experimental data remarkably well for both
$^{40}$Ca+$^{90}$Zr and $^{40}$Ca+$^{96}$Zr. Neither adjusting
parameters nor adding some special reaction channels for
neutron-rich nuclei(see ref.\cite{Deni98}) are needed in our
calculations. It implies that our improved QMD model is quite
successful in describing the fusion reaction near barrier for both
nuclei at $\beta$ stable line and neutron-rich nuclei.
In order to investigate the effect of neutron-rich projectile on
fusion probability in Fig.9 we show the
fusion cross sections of $^{48}$Ca+$^{90}$Zr.
One can see a even stronger enhancement of fusion cross sections
in this case.
\[
\fbox{Fig. 8 }
\]
\[
\fbox{Fig. 9 }
\]

From the comparison of the fusion cross sections for
$^{40}$Ca+$^{90}$Zr ,$^{40}$Ca+$^{96}$Zr
and $^{48}$Ca+$^{90}$Zr, one can easily find that there is a strong enhancement
of fusion cross section for neutron-rich nuclear reactions.
In order to study the mechanism of the enhancement of fusion cross sections
for neutron-rich nuclear reactions, we study the height of
dynamic Coulomb barrier $V_{b}$, the potential well of compound
nuclei $V_{com.}$ and as well as the neutron and proton density
distribution of the compound nuclei. Here the height of dynamic
Coulomb barrier means the height of the highest Coulomb barrier
experienced during the path of fusion. The method of calculating
dynamic Coulomb barrier is given in section 3.2. The potential
well of compound nuclei can be calculated by the following
expression:
\begin{equation}  \label{22}
V_{com.}({\bf r}^{\prime })=\int \rho _{com.}({\bf r})V({\bf
r-r}^{\prime })d^{3}{\bf r},
\end{equation}
where $\rho_{com.}({\bf r})$ is the density distribution of the
compound nuclei formed in the fusion reaction, and $V({\bf r-r}')$
is effective nucleon-nucleon interaction. The calculation results
for $^{40}$Ca+$^{90}$Zr ,$^{40}$Ca+$^{96}$Zr and
$^{48}$Ca+$^{90}$Zr at $E_{c.m.}=95.0MeV, 107.6MeV$ are listed in
Table.3.

Now, let us first discuss the effect of neutron-rich target by comparing reactions
$^{40}$Ca+$^{90}$Zr and $^{40}$Ca+$^{96}$Zr at energy near barrier. At
$E_{c.m.}=107.6MeV$, the average height of dynamic Coulomb barrier is
about $88.2MeV$ which is lower than that of static Coulomb barrier
of about $98.0MeV$, and the average of depth of mean potential well of
compound nuclei $V_{com.}$ is about $45.38MeV$. Comparing with
$^{40}$Ca+$^{90}$Zr case, one can see that at this energy the height of dynamic
Coulomb barrier is almost equal, and the depth of mean potential
well of compound nuclei formed in $^{40}$Ca+$^{96}$Zr is only about
$0.3MeV$ deeper than that in $^{40}$Ca+$^{90}$Zr which is about
$44.92MeV$. But as energy decreases, for
example, $E_{c.m.}=95.0MeV$, the height of dynamic Coulomb barrier
for $^{40}$Ca+$^{96}$Zr is about $80.6MeV$ and
is more than $5MeV$ lower than that of $85.2MeV$ for $^{40}$Ca+$^{90}$Zr.
While the depth of mean potential well of compound
nuclei formed in the fusion process at this energy increases a little
comparing with that at $E_{c.m.}=107.6MeV$ for both $^{40}$Ca+$^{90}$Zr
and $^{40}$Ca+$^{96}$Zr.
From this result, we find
that the dynamic Coulomb barrier for reactions with neutron-rich target such as
$^{40}$Ca+$^{96}$Zr decreases more strongly than that for $^{40}$Ca+$^{90}$Zr
as energy decreases
from 107.6 Mev to 95.0 MeV and consequently, it leads to a stronger
enhancement of the fusion cross sections for $^{40}$Ca+$^{96}$Zr at lower energy.

\[
\fbox{Fig. 10 }
\]

Now, let us turn to the fusion reaction with neutron-rich
projectile, $^{48}$Ca+$^{90}$Zr. Fig.10 shows the comparison
between the static and dynamic Coulomb barrier of
$^{48}$Ca+$^{90}$Zr at energy $E_{c.m.}=95.0MeV$. The solid curve
denotes the static Coulomb barrier, and the dotted curve denotes
the dynamic Coulomb barrier. From Fig.10, one can see that in the
fusion reaction at near barrier, the dynamic Coulomb barrier is
lower than the static Coulomb barrier and the thickness of the
barrier decreases largely. Concerning the height of dynamic
Coulomb barrier, from Table 3 one can see at energy
$E_{c.m.}=107.6MeV$, the height of its dynamic Coulomb barrier is
about $85.4MeV$ which is about 3 MeV lower than that for both
$^{40}$Ca+$^{90}$Zr and $^{40}$Ca+$^{96}$Zr cases. And at
$E_{c.m.}=95.0MeV$, the height of barrier falls about one MeV and
is lower than that of $^{40}$Ca+$^{90}$Zr but higher than that of
$^{40}$Ca+$^{96}$Zr. The depth of mean potential well of the
compound nuclei formed in $^{48}$Ca+$^{90}$Zr is a little deeper
than that formed in $^{40}$Ca+$^{90}$Zr. On the other hand, the
shape of the mean potential well formed in fusion process may also
play an important role for fusion probability. In Fig.11 we show
the mean potential wells of compound nuclei formed in
$^{40}$Ca+$^{90}$Zr(the dotted curve), $^{40}$Ca+$^{96}$Zr(the
dashed curve) and $^{48}$Ca+$^{90}$Zr(the solid curve). From the
comparison, one can see that the mean potential well of compound
system in $^{48}$Ca+$^{90}$Zr is obviously wider than the other
two cases. And when energy decreases from $E_{c.m.}=107.6MeV$ to
$E_{c.m.}=95.0MeV$, the depth of mean potential well of compound
nuclei in $^{48}$Ca+$^{90}$Zr increases more than the other two
cases(see Table.3). In order to understand the reason of forming
the different mean potential well for these three different
reactions. We show the neutron and proton density distributions of
compound nuclei for three cases in Fig.12 in which the dashed
curves denote the density distribution of neutrons and the solid
curves denote that of protons. From Fig.12 one can see that the
proton distribution for reaction $^{48}$Ca+$^{90}$Zr is different
from the other two cases and also there are relatively more
neutrons distributed on the surface of compound nuclei.  This kind
of density distribution seem to have advantage of forming a
favorable potential well and leads to a strong enhancement of
fusion probability for $^{48}$Ca+$^{90}$Zr.\\

\[
\fbox{Table. 3}
\]

\[
\fbox{Fig. 11 }
\]
\[
\fbox{Fig. 12 }
\]

From above comparison of the fusion cross sections for reactions
$^{40}$Ca+$^{90}$Zr, $^{40}$Ca+$^{96}$Zr and $^{48}$Ca+$^{90}$Zr.
We find that there is an enhancement of fusion cross sections for
reactions with neutron-rich target or projectile resulting from
the lowering of dynamic Coulomb barrier or forming a favorable
potential well of compound system in fusion process or both.
However, as soon as two nuclei approach with each other their
shape could be deformed and after contact a neck will be
developed. The fusion cross section can be affected largely by
this dynamical process. What is the role played by excess neutrons
in neutron-rich projectile and target on this dynamical process?
How do protons and neutrons transfer during fusion process? All
these problems are very important for understanding the mechanism
of fusion reactions. They will be studied in our future work.
\begin{center}
{\bf CONCLUSION}
\end{center}

In this work, we have proposed an improved QMD model. Our
improvements mainly include: taking into account the effects of
surface term; introducing system size dependent wave packet width;
and adopting phase space constraint of  \={f}$_{i}\leq 1$. By
using this model, the ground state properties including binding
energy, mean square root of radius, density distribution, momentum
distribution and phase space distribution and so on from $^{6}$Li
to $^{208}$Pb can be described very well with one set of
parameters. Simultaneously the Coulomb barrier can be described
well. By applying our improved QMD model, the experimental data of
the fusion cross sections for $^{40}$Ca+$^{90,96}$Zr can be
reproduced remarkably well without introducing any new parameters.
In addition, the fusion reaction at energy near barrier for
$^{48}$Ca+$^{90}$Zr is studied. We find that there is an
enhancement of fusion cross sections for reactions with
neutron-rich target or projectile. The mechanism for the
enhancement of fusion cross sections in these reactions is the
lowering of the dynamic Coulomb barrier or forming the favorable
potential well of compound system in fusion process or both.
Nevertheless, the problems concerning the neck dynamics and the
mass transformation are not the task of this paper. The work about
these aspects is on progress.

\newpage

\begin{center}
{\small {\bf CAPTIONS} }
\end{center}

\begin{description}
\item[{\tt Fig.1}]  The schematic figure of the effect of surface term of interaction potential energy.
(a) the density distribution of Boltzmann form. (b) the shape of
interaction of surface term, the arrows denote the direction of
corresponding force. (c) the comparison between the density distribution calculated
with and without surface term taken into account.
The solid curve denotes the density distribution with surface
term, and the dashed curve denotes that of without surface term.
\item[{\tt Fig.2}] The time evolution of density distribution of
$^{90}$Zr without surface term.
\item[{\tt Fig.3}] The time evolution of density distribution of
$^{90}$Zr with surface term taken into account.
\item[{\tt Fig.4}] The time evolution of momentum distribution of $^{208}$Pb.
The dash-dotted curve denotes the initial momentum
distribution which is obtained from the Relativistic Mean Field
calculations. The dashed curve denotes the momentum distribution
without phase space density constraint taken into account.
The solid curve denotes the momentum distribution with constraint.
\item[{\tt Fig.5}] The time evolution of binding energies and mean
square roots of radii of $^{16}$O,$^{40}$Ca,$^{90}$Zr and $^{208}$Pb.
\item[{\tt Fig.6}] The static Coulomb barrier of $^{40}$Ca+$^{90}$Zr
calculated by QMD model and proximity potential. The
solid curve denotes the results of QMD model calculation and the crossed curve denotes
that of proximity potential, respectively.
\item[{\tt Fig.7}] The comparison between the static Coulomb barrier of
$^{40}$Ca+$^{90}$Zr calculated with fixed wave packet width (the
dashed curve) and the system size dependent wave packet width(the
solid curve).
\item[{\tt Fig.8}] The fusion cross sections for $^{40}$Ca+$^{90,96}$Zr. The experimental data
are taken from\cite{Timm98}.
 The solid curves denote the results of QMD model and the
crossed curves denote the experimental data.
\item[{\tt Fig.9}] The fusion cross section for $^{48}$Ca+$^{90}$Zr.
\item[{\tt Fig.10}] The Coulomb barrier for fusion reaction
$^{48}$Ca+$^{90}$Zr. The solid curve denotes the static Coulomb
barrier and the solid curve with dots denotes the dynamic Coulomb barrier at
energy $E_{c.m.}=95.0MeV$.
\item[{\tt Fig.11}] The comparison of mean potential wells of compound nuclei
formed in $^{40}$Ca+$^{90}$Zr, $^{40}$Ca+$^{96}$Zr and in
$^{48}$Ca+$^{90}$Zr. The dotted curve denotes the mean potential
well of compound nuclei in fusion reaction $^{40}$Ca+$^{90}$Zr,
the dashed curve denotes that in reaction $^{40}$Ca+$^{96}$Zr and the
solid curve denotes that in reaction $^{48}$Ca+$^{90}$Zr.
\item[{\tt Fig.12}] The density distributions of compound nuclei formed in
$^{40}$Ca+$^{90}$Zr, $^{40}$Ca+$^{96}$Zr and
$^{48}$Ca+$^{90}$Zr. The dashed curves denote the neutron density
distribution and the solid curves denote proton density distribution.
\item[{\tt Table.1}] The parameters adopted in the present work.
\item[{\tt Table.2}] The calculating results of binding energies per
nucleon and mean square roots of radii for nuclei from $^{6}$Li to $^{208}$Pb. The binding
energies are compared with the experimental data. The mean square
roots of radii are compared with empirical formula\cite{Pre75}.
\item[{\tt Table.3}] The height of dynamic Coulomb barrier and the
depth of mean potential wells of compound nuclei in fusion reactions
$^{40}$Ca+$^{90}$Zr, $^{40}$Ca+$^{96}$Zr and $^{48}$Ca+$^{90}$Zr
at $E_{c.m.}=95.0MeV$ and $E_{c.m.}=107.6MeV$.
\end{description}

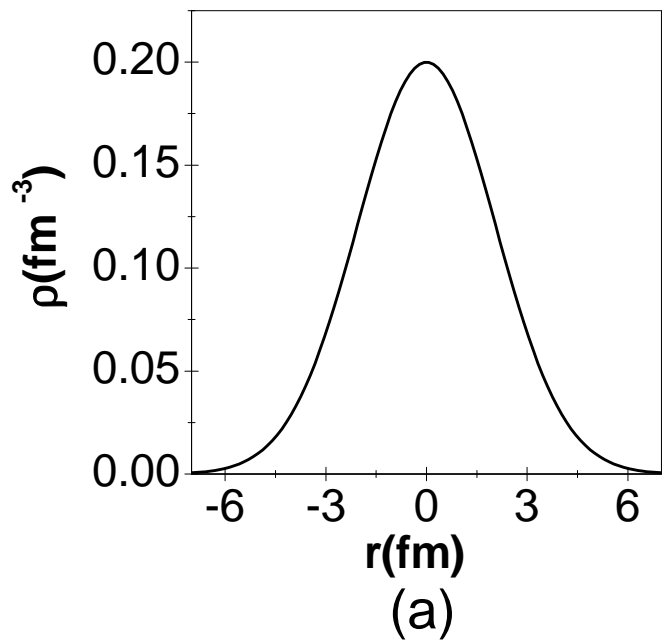
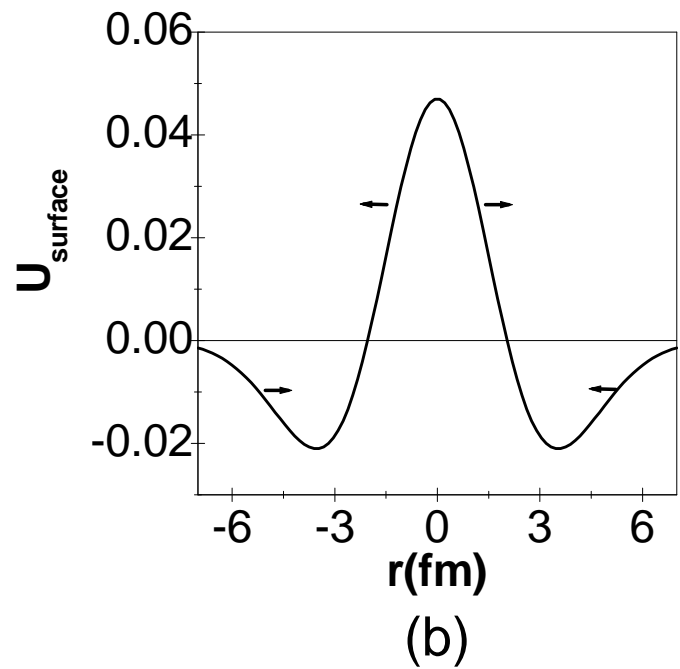
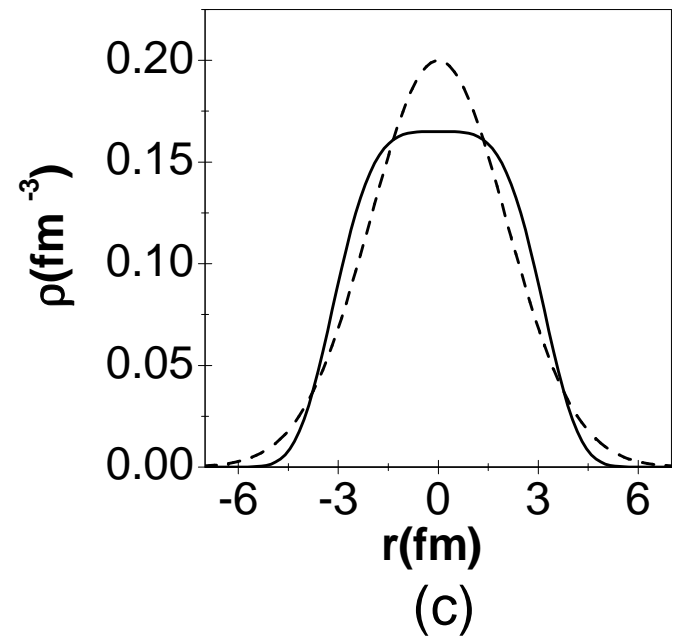
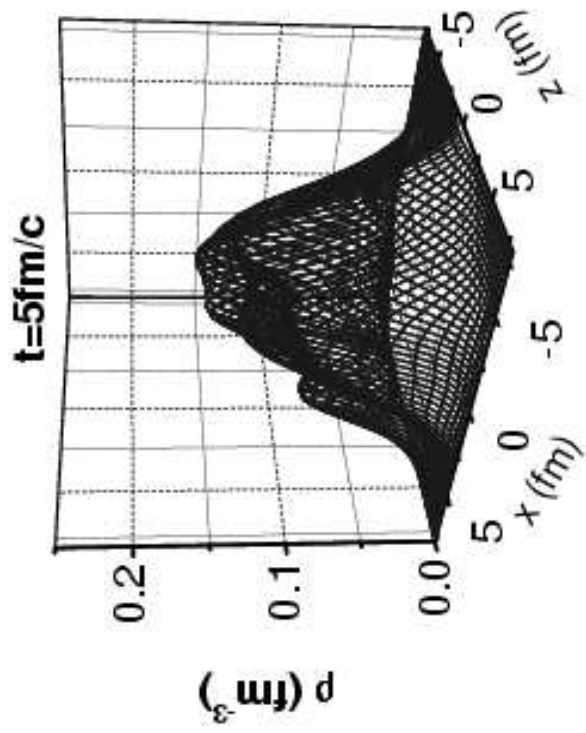
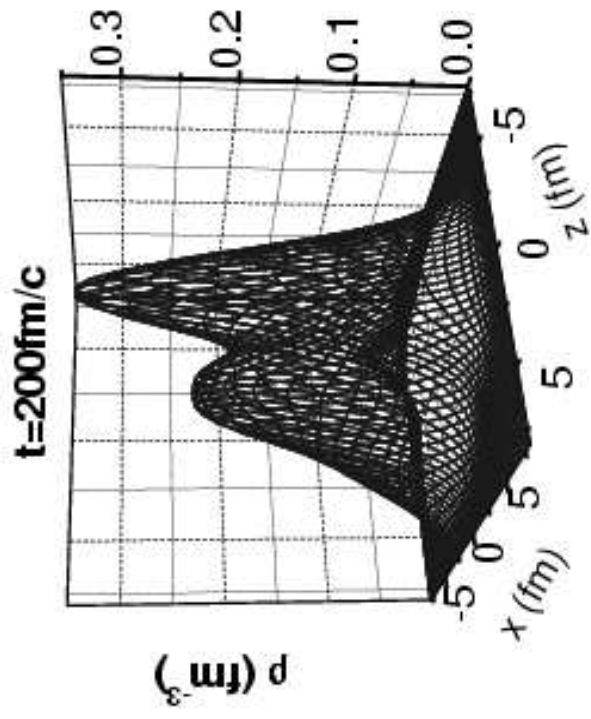
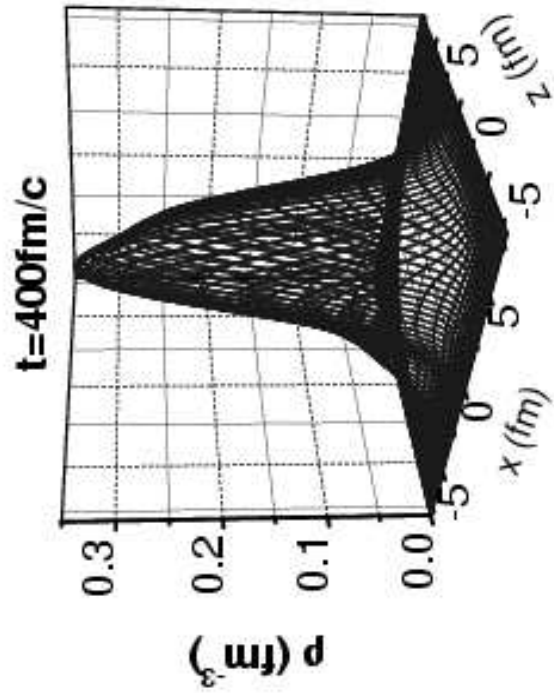
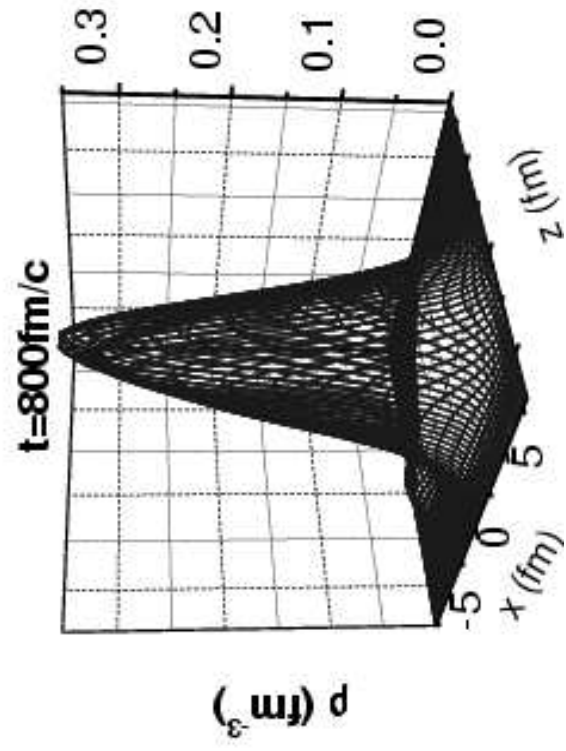
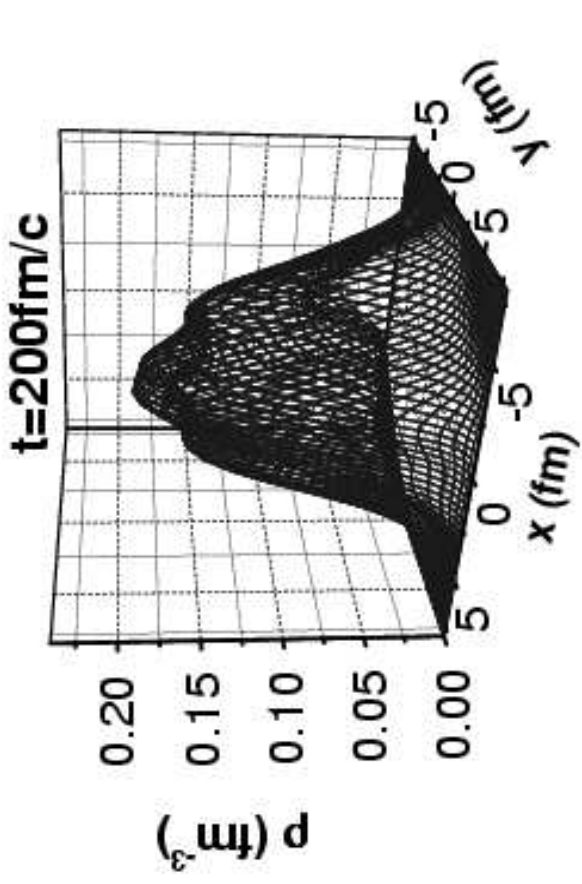
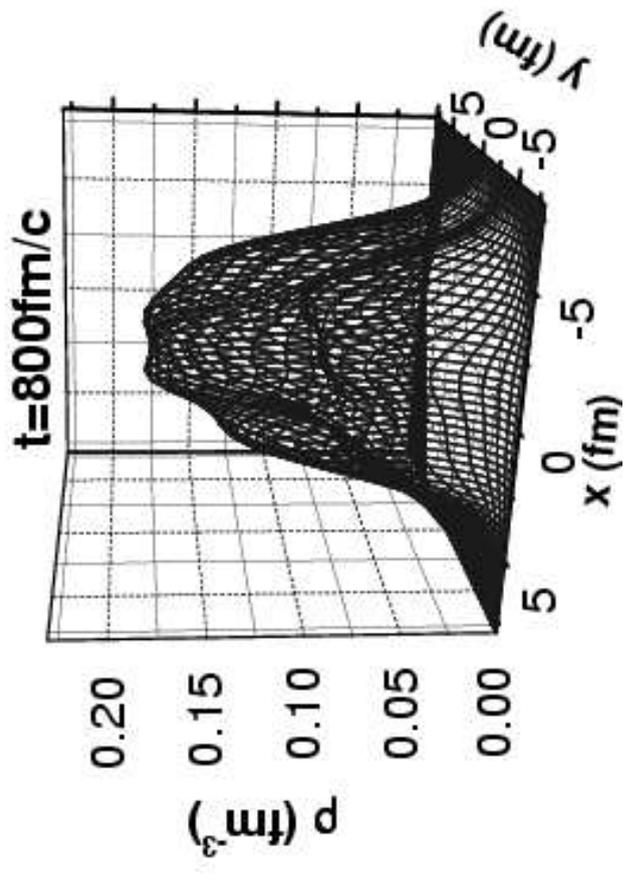
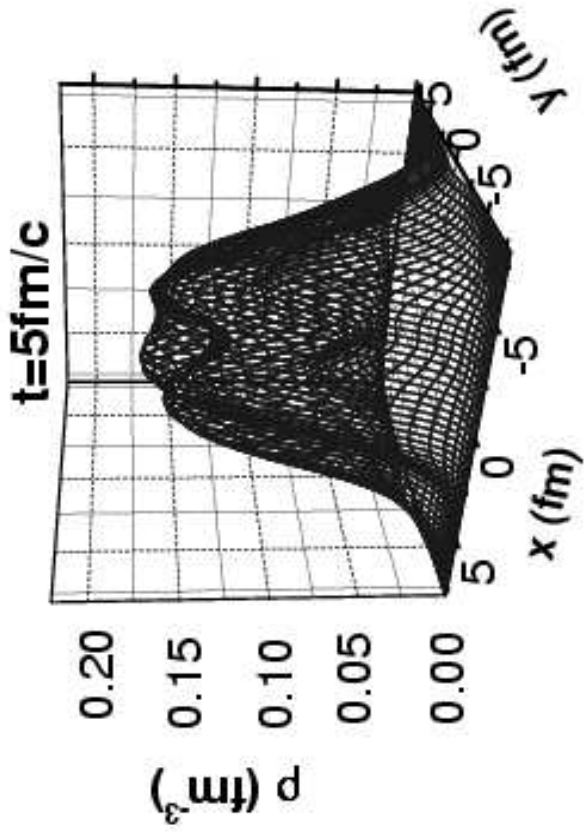
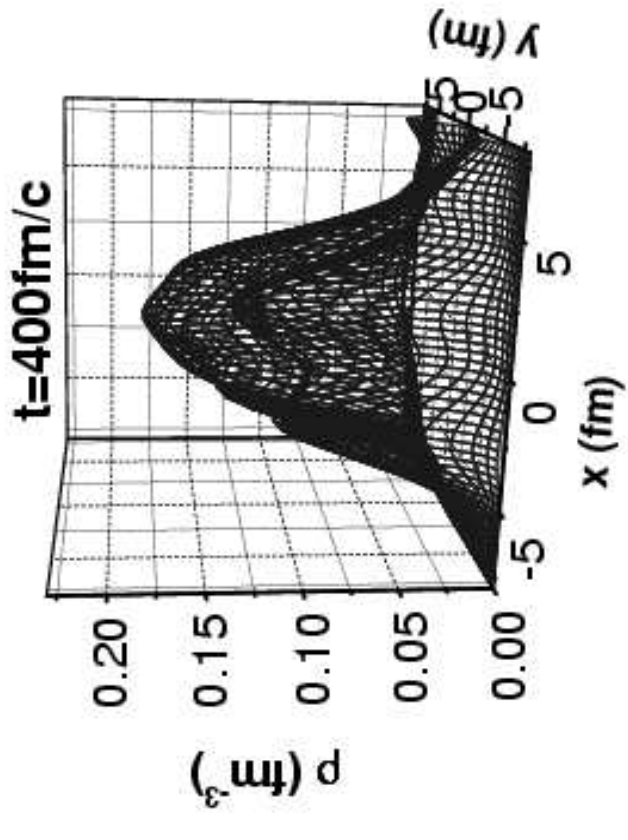
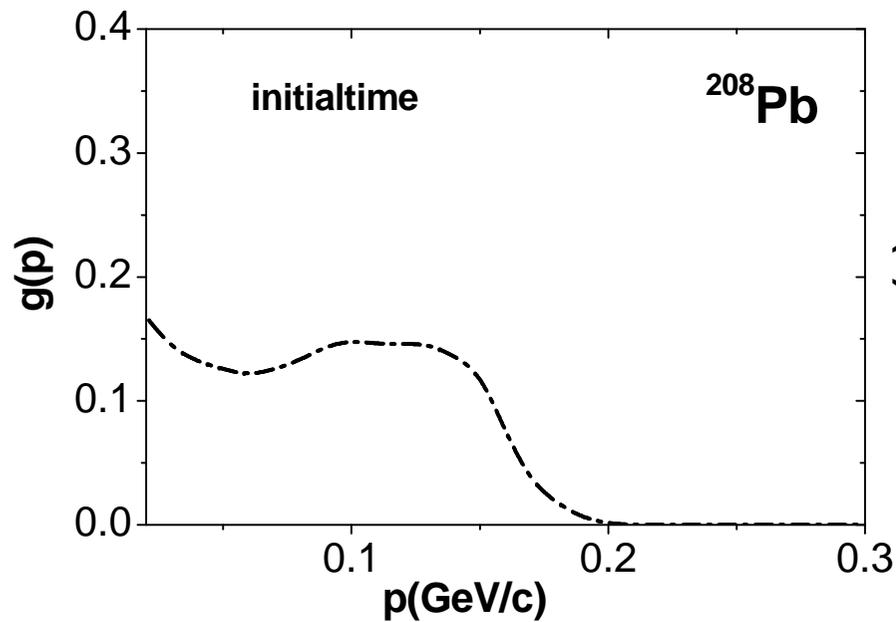
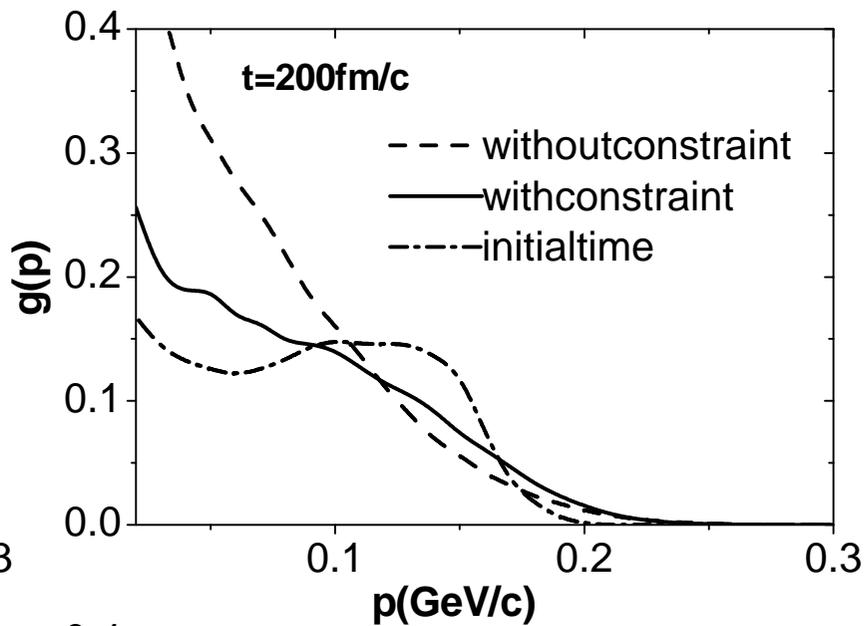
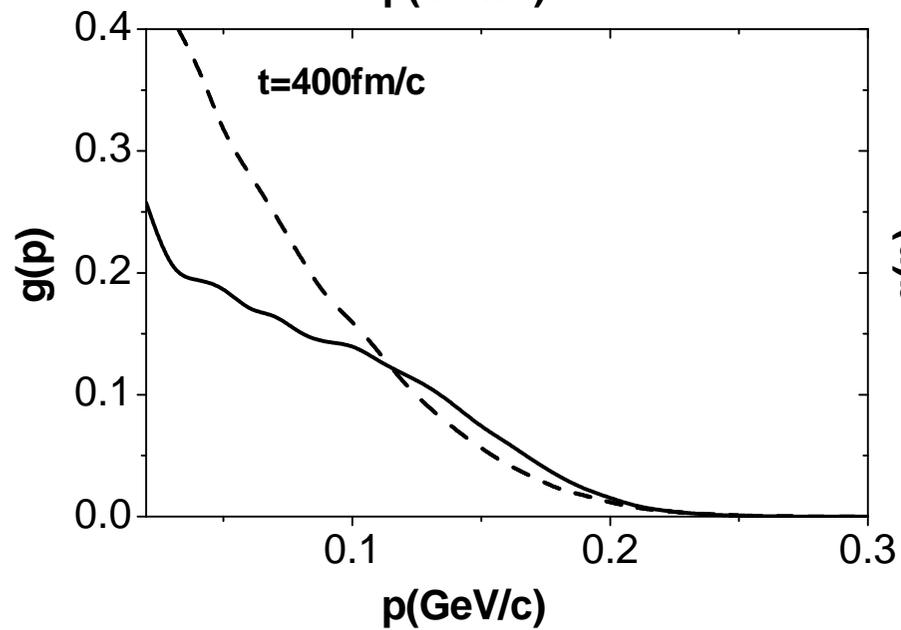
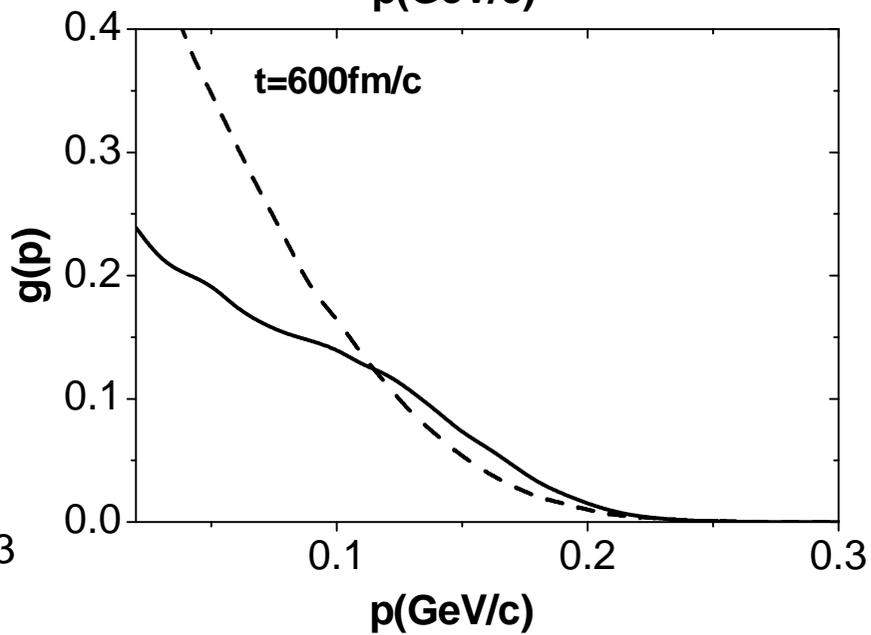
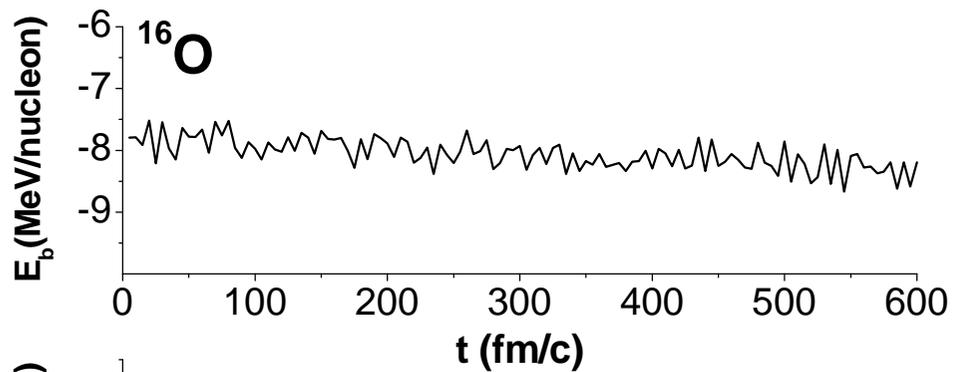
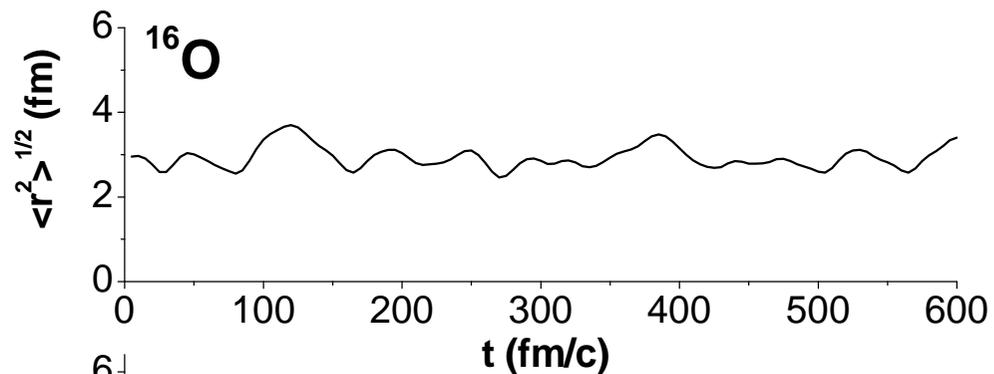
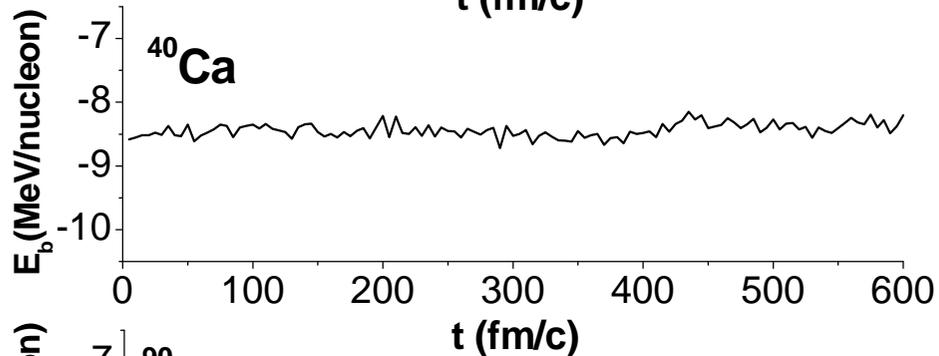
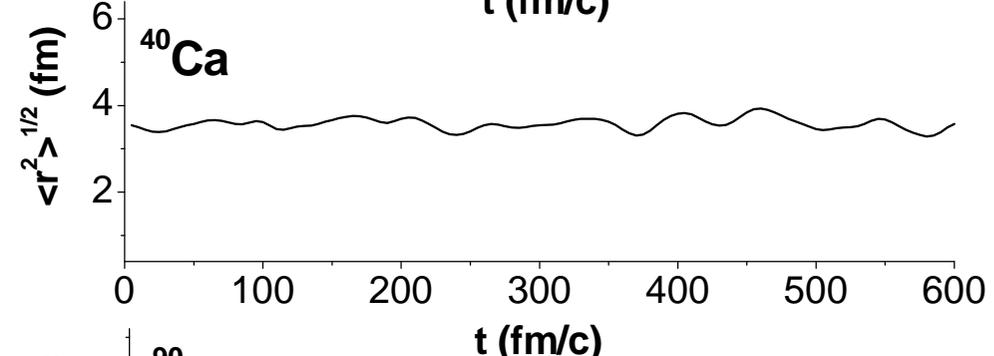
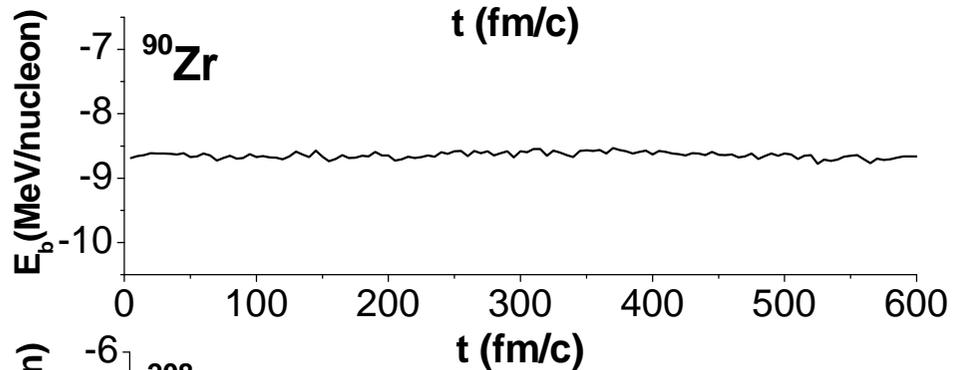
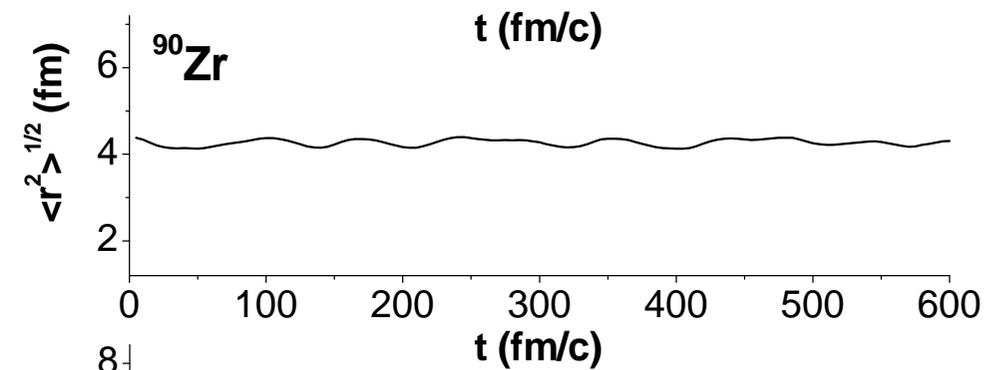
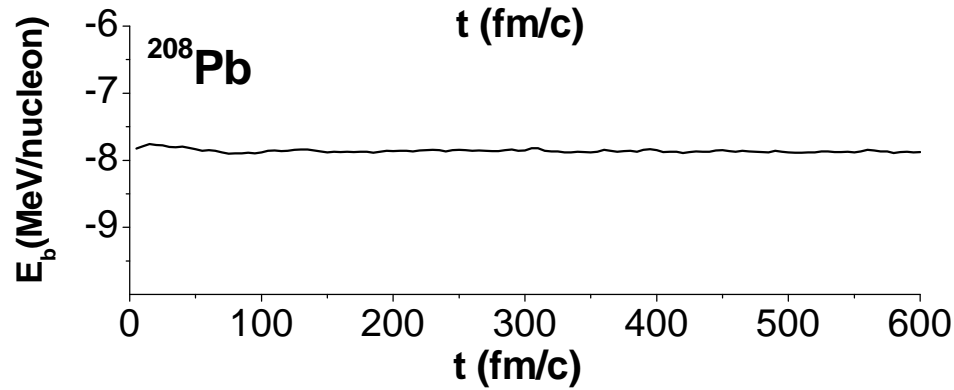
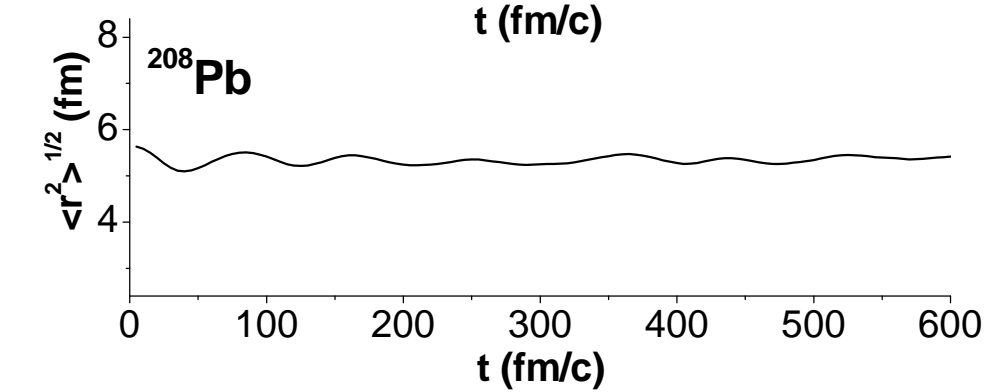
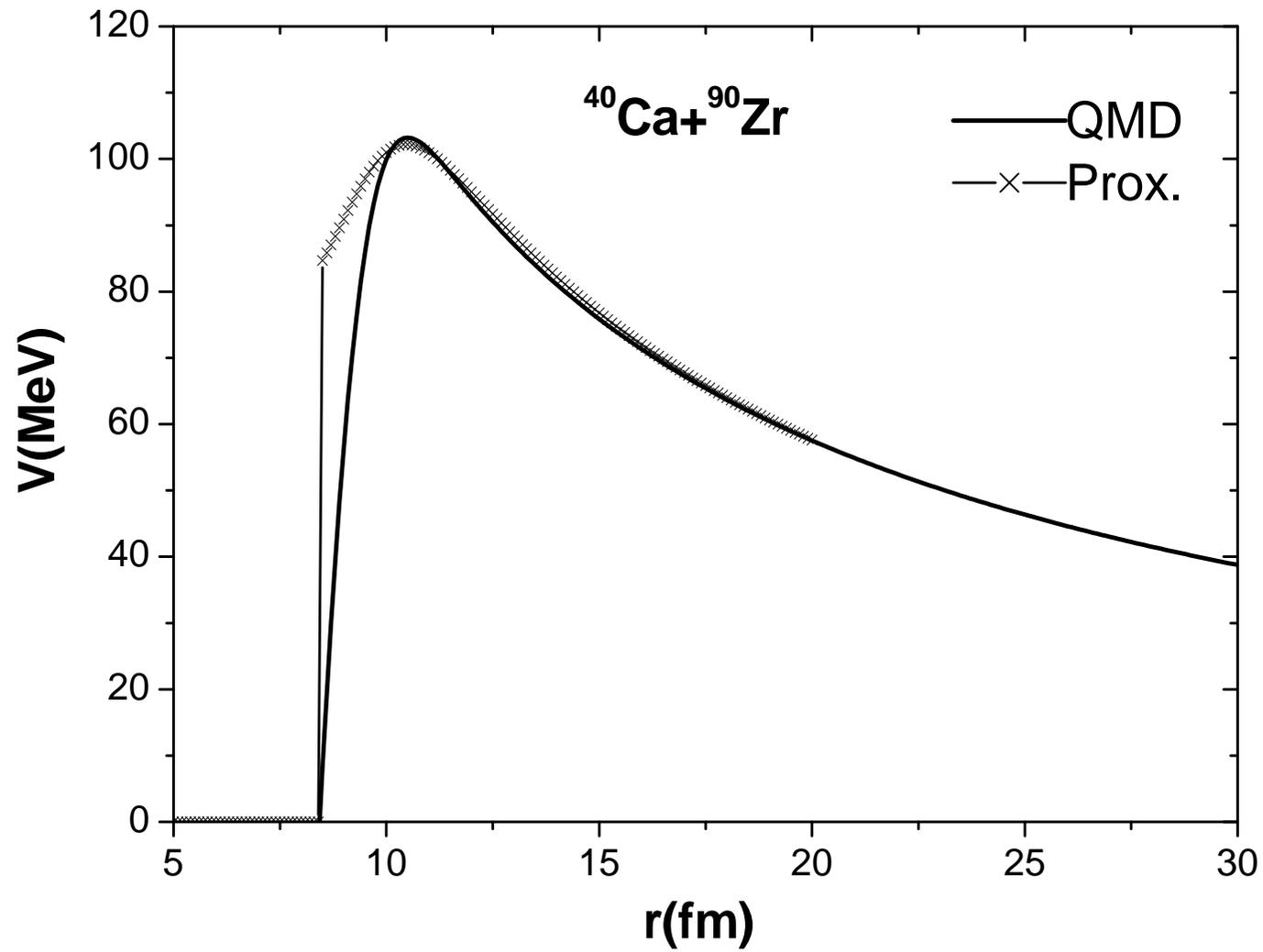
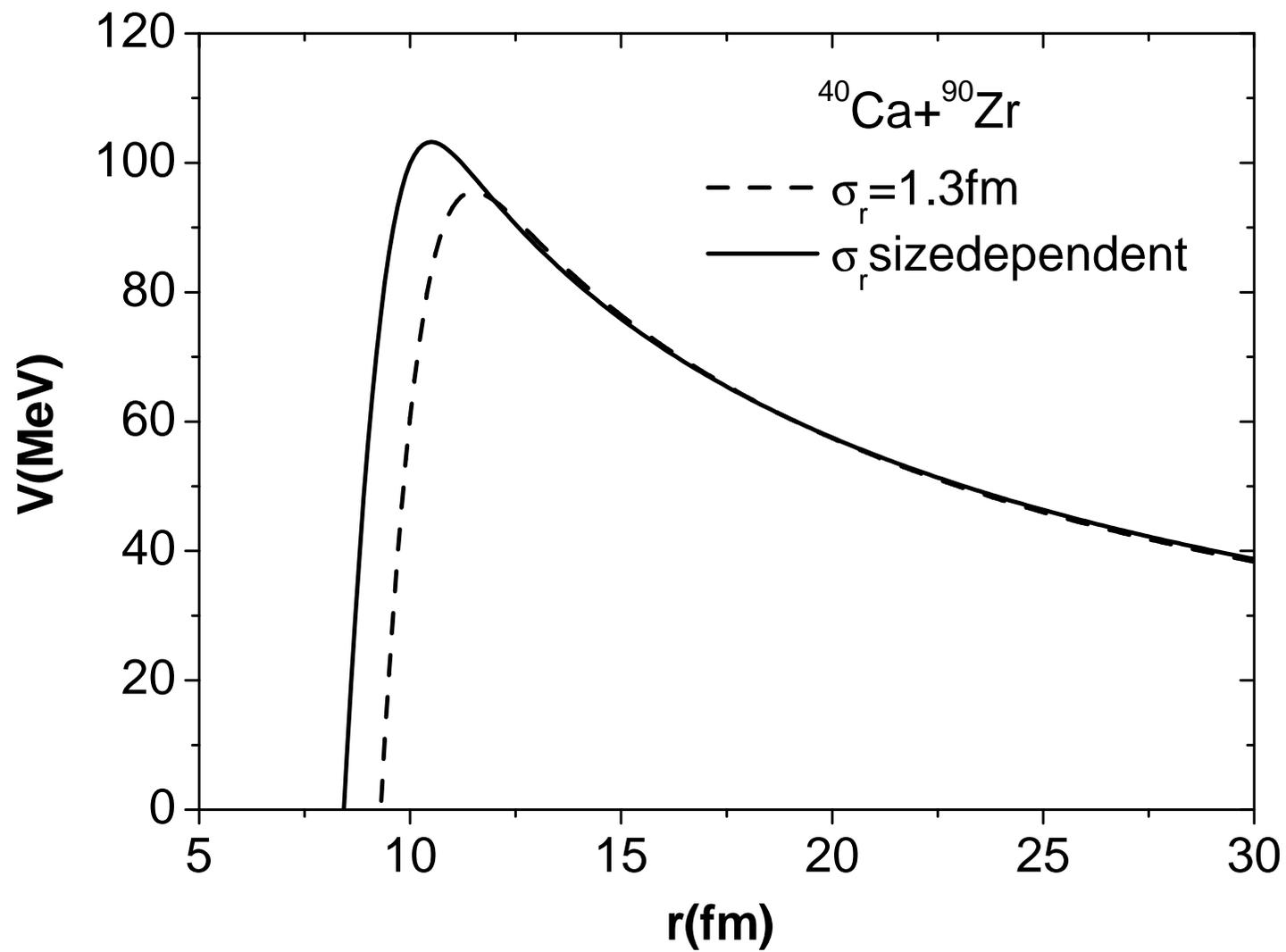
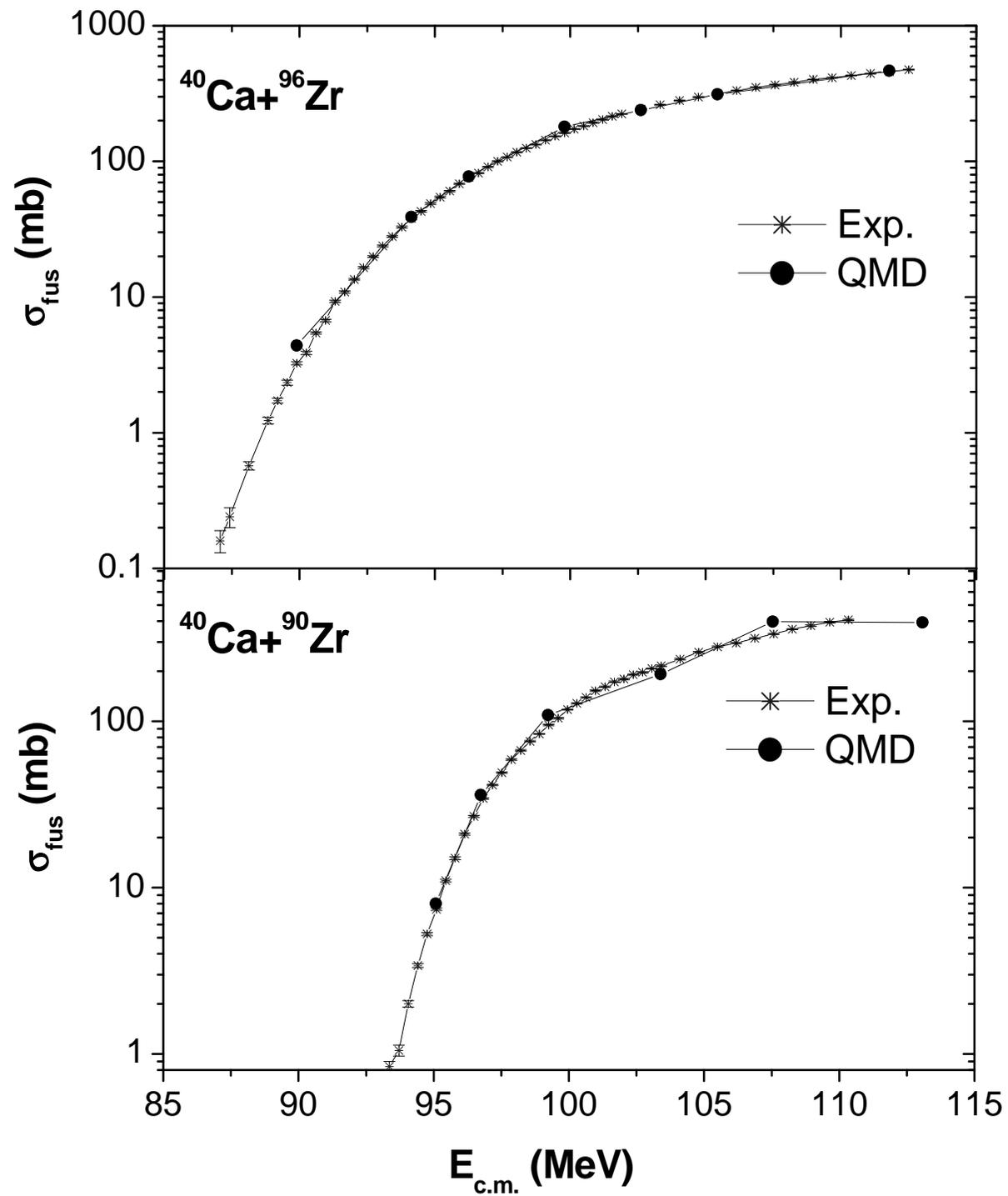
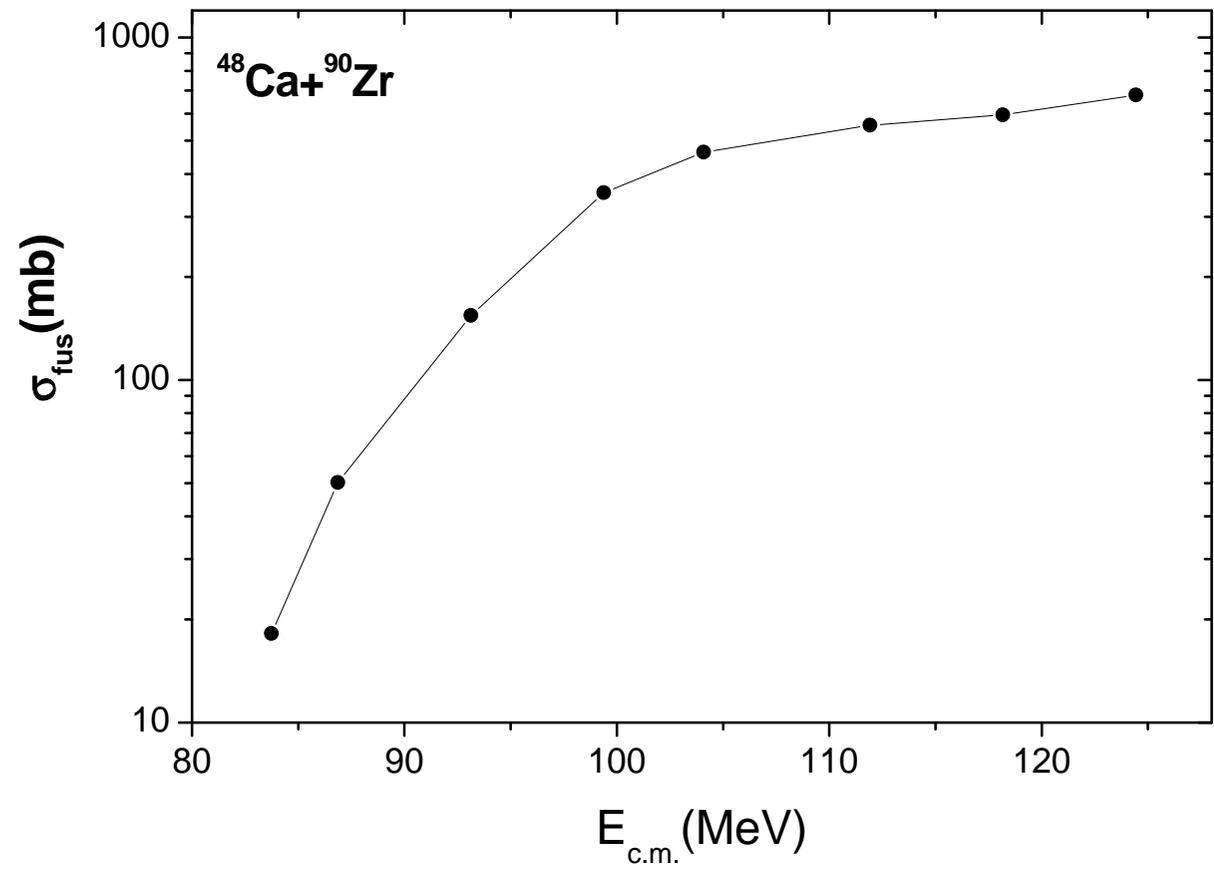
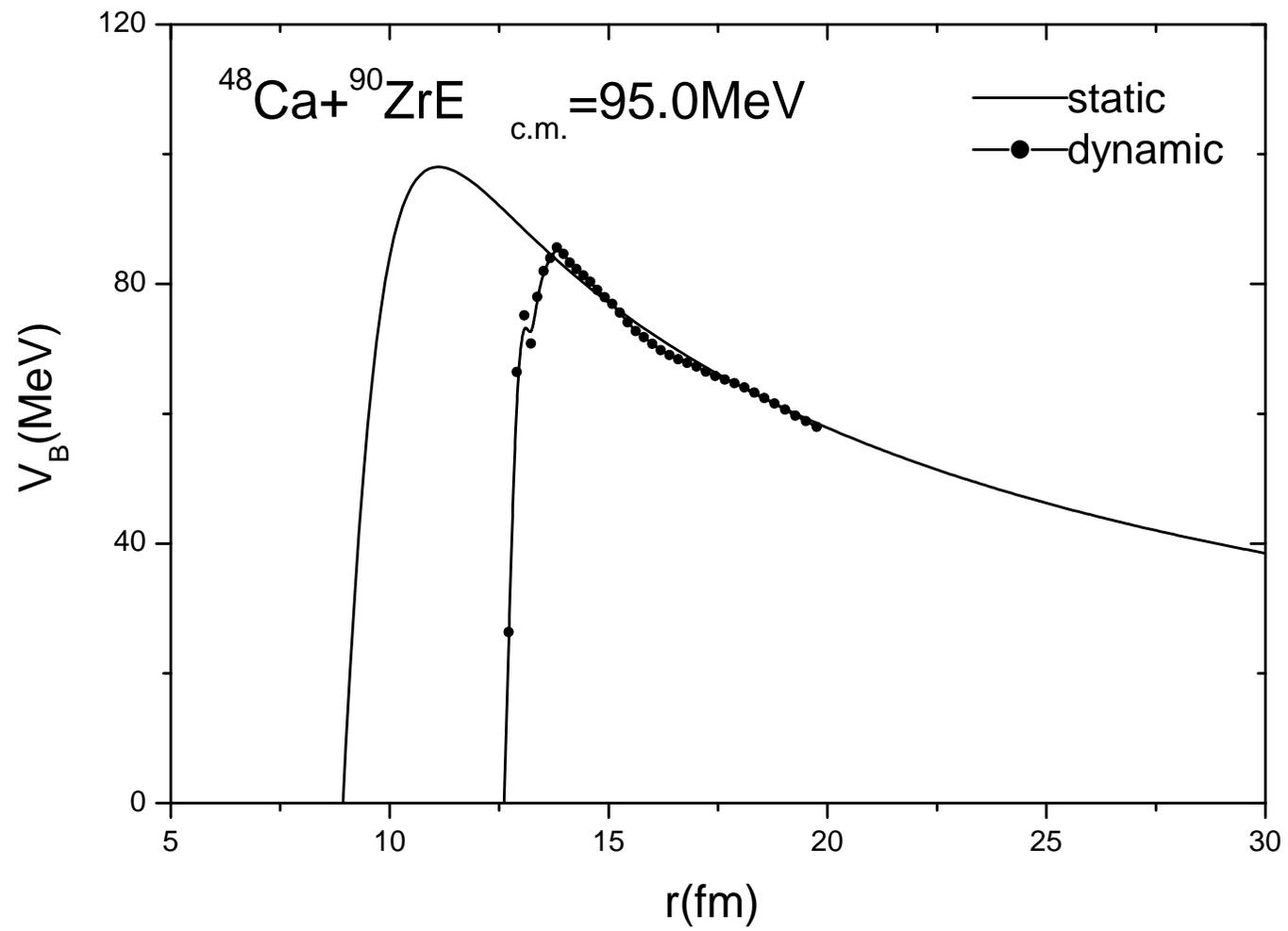
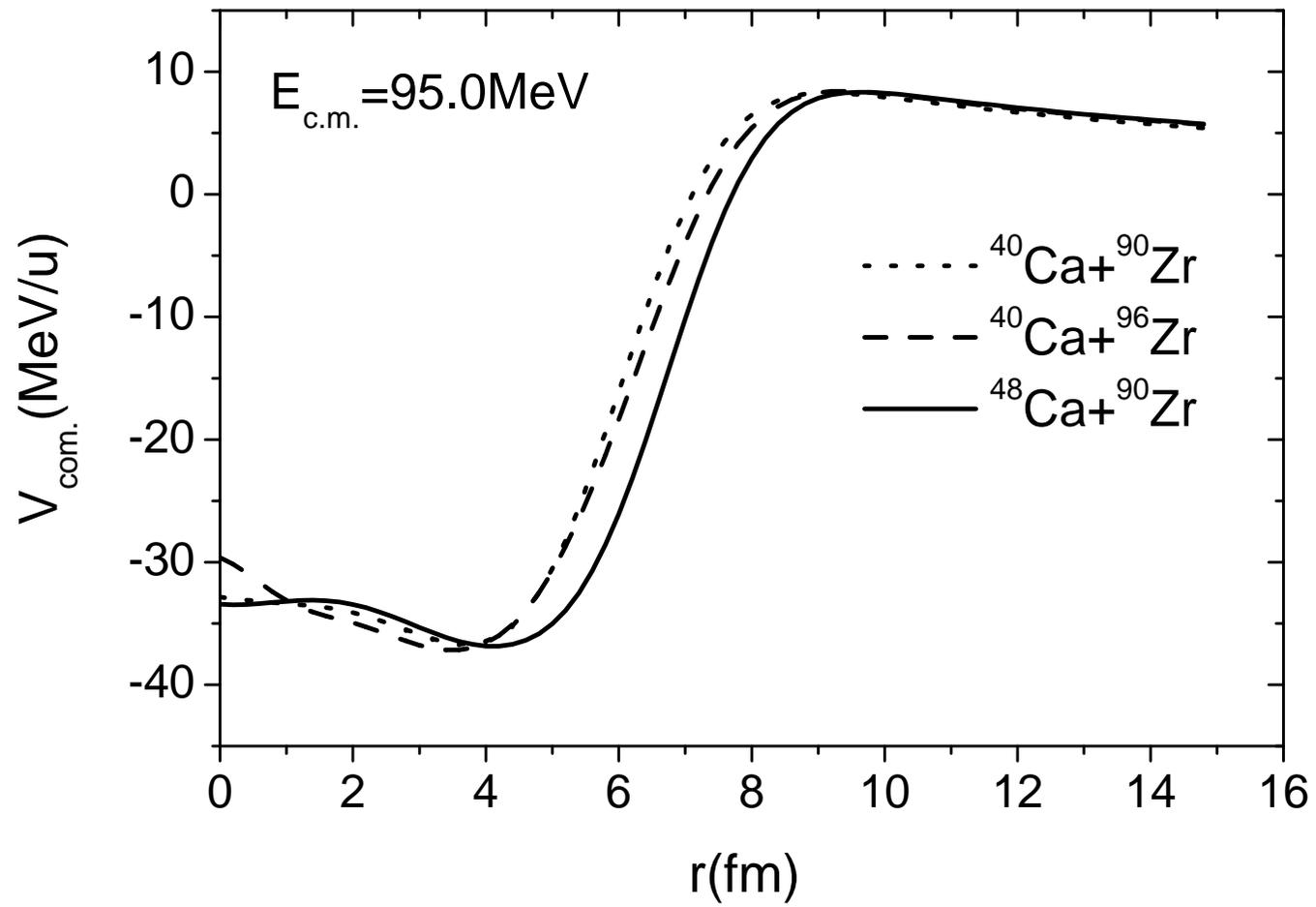
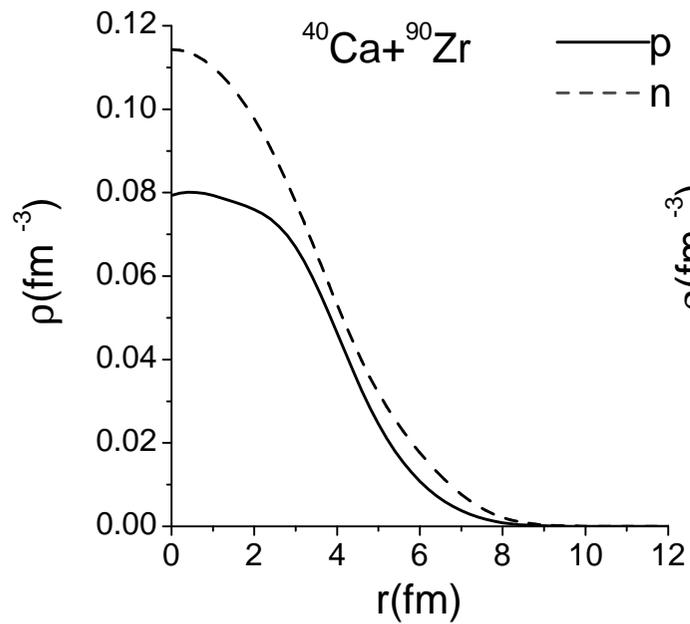
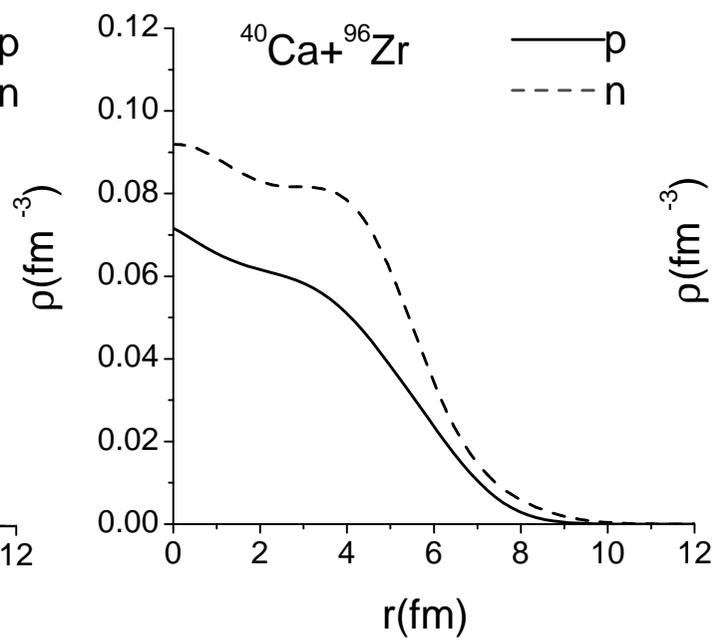
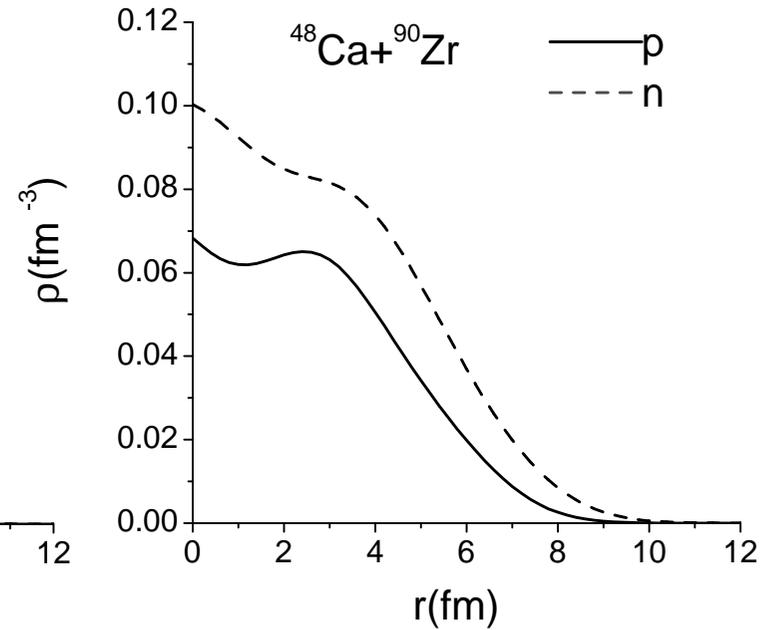
\end{document}